\documentclass[preprint]{elsarticle}

\usepackage{graphicx}
\usepackage{epsfig}
\usepackage{amsmath}
\usepackage{tensor}

\usepackage{xcolor}
\usepackage{amsfonts}
\usepackage[normalem]{ulem}
\usepackage{graphicx}
\usepackage{subcaption}
\usepackage{amssymb}
\usepackage{hyperref}
\usepackage{cleveref}
\usepackage{placeins}

\usepackage{colortbl}
\usepackage{mathtools}
\usepackage{mathbbol}
\usepackage{eucal}

\usepackage{colortbl}
\usepackage{mathtools}
\usepackage{mathbbol}
\usepackage{eucal}
\usepackage{xcolor}

\newcommand{\be}{\begin{align}}
\newcommand{\ee}{\end{align}}
\newcommand{\bea}{\begin{eqnarray}}
\newcommand{\eea}{\end{eqnarray}}

\newcommand{\nd}{\noindent}

\crefname{equation}{Eq.}{Eqs.}
\Crefname{equation}{Equation}{Equations}

\begin{document}

\title{Spatiotemporal Moran dynamics in continuous media}

\author[1,2]{Melika Gorgi\corref{cor1}\fnref{eq1}}
\author[3] {Kamran Kaveh\fnref{eq1,eq2}}
\author[4]{Navid Aliakbarian}
\author[4]{Mohammad Reza Ejtehadi}

\cortext[cor1]{Corresponding author: mgorgi@uci.edu}

\fntext[eq1]{These authors contributed equally to this work.}
\fntext[eq2]{Affiliation when this work was conducted.}

\address[1]{Center for Complex Biological Systems, University of California, Irvine, CA 92697, USA.}
\address[2]{Department of Physics and Astronomy, University of Southern California, Los Angeles,CA 90089, USA.}
\address[3]{Therapy Modeling and Design Center, University of Minnesota, Minneapolis,MN 55455 , USA.}
\address[4]{Department of Physics, Sharif University of Technology, Tehran, Iran}

\begin{abstract}
 Understanding how natural selection unfolds across space and time is a central problem in evolutionary biology. Classical models such as the Moran process capture stochastic birth–death dynamics in structured populations, while reaction–diffusion equations like the Fisher–Kolmogorov–Petrovsky–Piskunov (FKPP) equation describe deterministic wavelike spread. In this work, we bridge these perspectives by deriving partial differential equations for the spatiotemporal limit of Moran dynamics in continuous media. Our model incorporates two distinct fitness components: fecundity (birth rate) and viability (death rate). We demonstrate that the resulting selective wave speeds differ substantially in spatial Moran birth-death (BD), Moran death-birth (DB), and FKPP dynamics. When fecundity drives the dynamics, we observe that the selective waves decelerate for the BD process, whereas in the DB process the wave propagates with a higher, constant speed. In contrast, when viability drives the process, the DB wave accelerates, while the BD and FKPP waves maintain comparable constant speeds. We extend the framework to heterogeneous media, represented as weighted lattice graphs in one or two dimensions. We derive a continuous-space analog of isothermal graphs and establish that the isothermality condition corresponds to the conservation of a local current.
\end{abstract}

\begin{keyword}
natural selection, spatial evolution, birth-death process, migration and motility, evolutionary graphs, isothermal theorem. 
\end{keyword}
\maketitle \vspace{10pt}

\section{Introduction}

\begin{figure}[h]
\begin{center}
\includegraphics[width=0.9\textwidth]{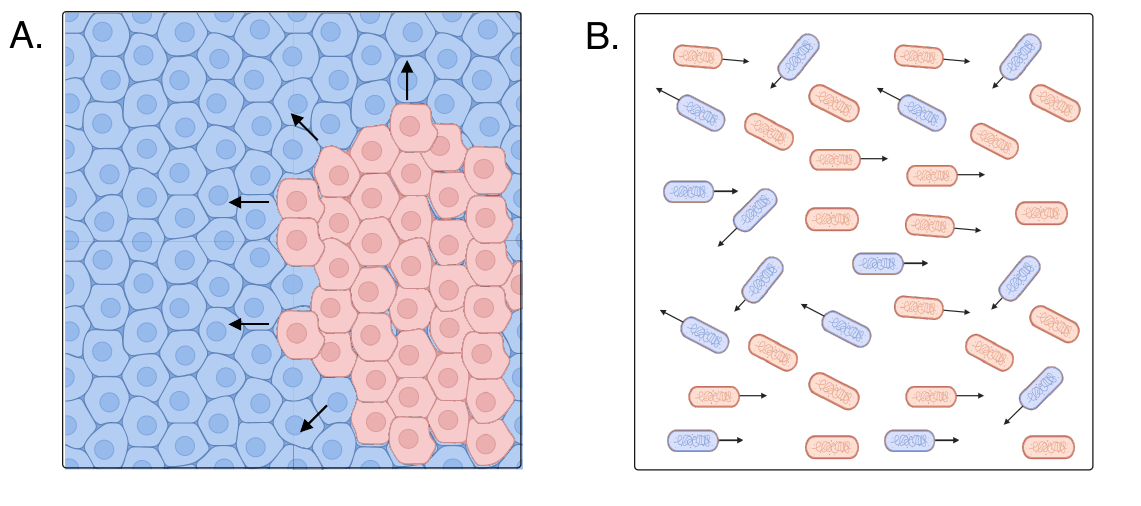}
\end{center}
    \caption{
            \textbf{Crowded vs dilute population modeling schemes in spatial evolutionary dynamics.}
            \textbf{(A)} A snapshot of a spatially structured population represented by a packed epithelial-like cell lattice. Each polygon denotes a single cell, colored red or blue to indicate mutant or resident genotypes, respectively. Mutants expand via local replacement events at boundaries between neighboring cells.
            \textbf{(B)} A dilute limit where individuals (cells) move randomly in continuous space and, independently, reproduce based on fitness differences, as modeled by FKPP reaction-diffusion equations.
            }
\label{fig1}
\end{figure}

Evolutionary dynamics describes how advantageous traits arise and spread within a population \cite{nowak2006evolutionary,moran1962,kimura1962probability,broom2014game,ewens2004mathematical,durrett2008probability}. Classical models of evolution often consider well-mixed populations, where spatial structures are neglected. However, the presence of structured populations and, in particular, spatial structures can change the outcome of an evolutionary process (\Cref{fig1}). Understanding evolutionary mechanisms in spatially structured populations is crucial in evolutionary biology, ecology, and applied fields such as epidemiology and cancer dynamics \cite{nowak2000virus, waclaw2015spatial,noble2022spatial,komarova2006spatial}. Measures of evolutionary success in a selection process, such as the chance of success of an advantageous allele, fixation probability, or its speed of spread, are profoundly different between a well-mixed population and a more realistic spatially structured population.

There are two approaches to study the dynamics of selection in spatial structures. First, agent-based models, such as Moran or Wright-Fisher, are useful in finite-population settings where stochasticity and drift are prominent. The spatial versions of Moran (or Wright-Fisher) models are defined on a lattice or a graph, which are known as evolutionary graphs. The nodes on the graph represent habitats, and the edges show the neighborhood and proximity between different habitats. Two common variations of the Moran dynamics are birth-death (BD) or death-birth (DB) update rules \cite{moran1962, nowak2006evolutionary,lieberman2005evolutionary,ohtsuki2006simple, kaveh2014duality}.

The second approach uses the deterministic models, often represented by a system of partial differential equations. They can describe the formation of mutant colonies over space-time and predict the speed of spread and topology of the mutant front. In particular, the Fisher-Kolmogorov-Petrovsky-Piskunov equation and its extension of the diffusion-reaction equation \cite{fisher1937wave, kolmogorov1937etude, murray2002mathematical,hallatschek2007genetic, korolev2010genetic} are widely used in diverse fields such as ecology, epidemiology, cancer modeling, and evolutionary genetics. (Throughout this paper, we refer to this model simply as `FK' for brevity.) It provides fundamental insights into spatial invasion and wavelike spread of beneficial alleles \cite{murray2002mathematical, van2003front, korolev2010genetic}. The FK equation, when applied to Darwinian selection, describes the competition between two genotypes, mutants (A) and residents (B), residing in a continuous medium. The mutant's (resident's) (reproductive) fitness is $r_{\rm A} (r_{\rm B})$, respectively. It is often written as a partial differential equation for the mutant frequency $\phi({\bf x},t)$: 

\begin{align}
\frac{\partial \phi({\bf x},t)}{\partial t} =  D\nabla^2 \phi({\bf x},t) + s \phi({\bf x})(1 - \phi({\bf x},t)).
\label{fk}
\end{align}
  
\begin{figure}[h]
\begin{center}
\includegraphics[width=1\textwidth]{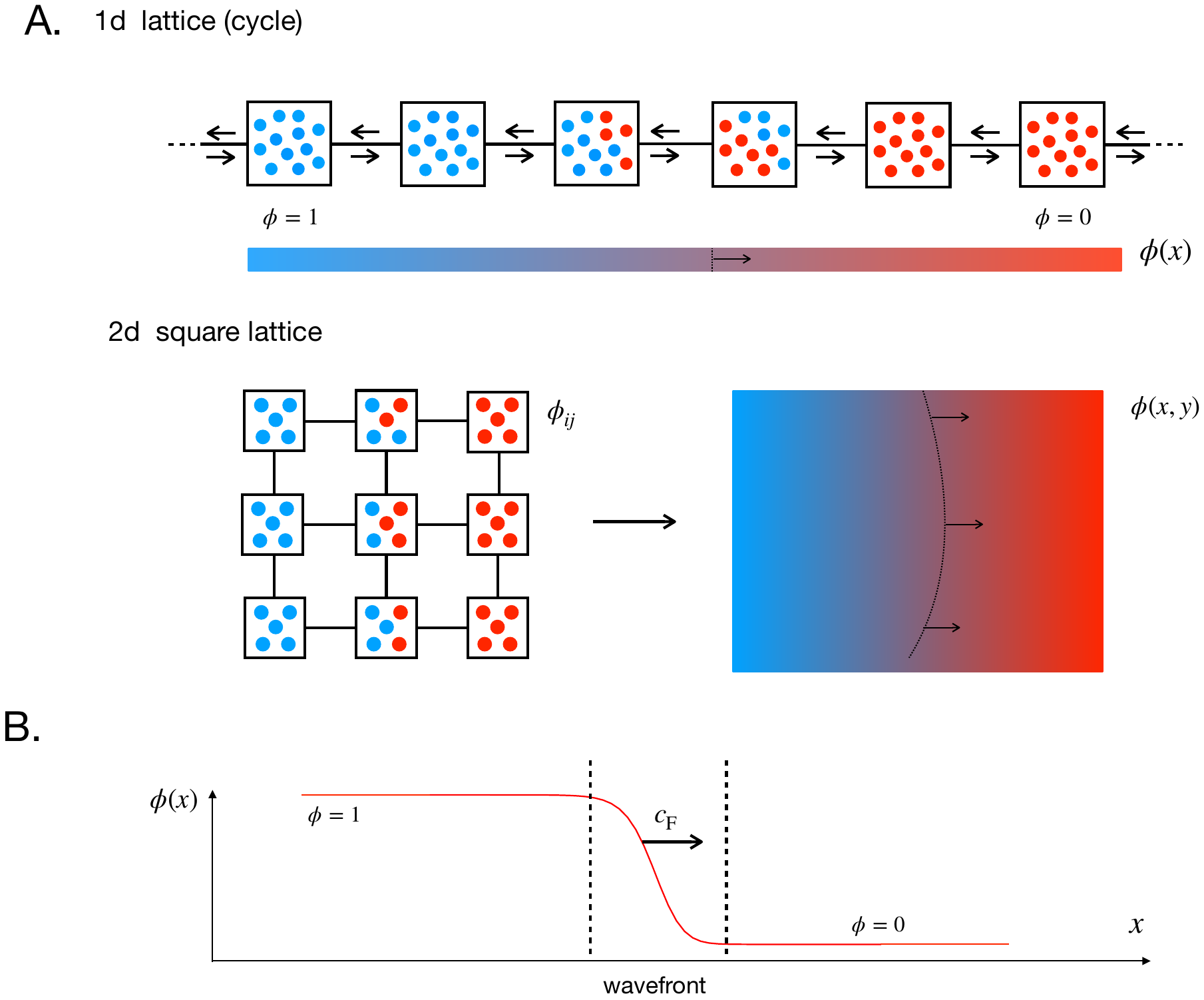}  
\end{center}
    \caption{
        \textbf{Lattice representations and traveling wave profile in spatial Moran dynamics.}
        \textbf{(A)} Schematic of discrete spatial structures used in modeling Moran dynamics. Top: a 1D cycle (ring) of demes (islands), each containing a fixed number of individuals. Mutant frequency \(\phi(x)\) varies across space, from \(\phi=1\) (blue) to \(\phi=0\) (red), forming a selective gradient. Bottom: a 2D square lattice representation where each node contains a well-mixed subpopulation with local mutant frequency \(\phi_{ij}\). The continuous approximation yields a mutant field \(\phi(x, y)\) with a spatial wavefront.
        \textbf{(B)} Illustration of the traveling wave solution \(\phi(x)\) in the continuum limit. The wavefront connects two homogeneous states, \(\phi=1\) and \(\phi=0\), and propagates with velocity \(c_F\). The width of the transition region corresponds to the diffusive spread of the advantageous allele.
    }
\label{fig2}
\end{figure}

\nd where the selection coefficient, $s = r_{\rm A} - r_{\rm B}$, is the difference between the fitness of two alleles. This equation combines diffusive dispersal with logistic growth and predicts traveling wave solutions characterized by a minimal invasion speed: $c_{\text{min}} = 2\sqrt{Ds}$. Similarly, the wavefront width, $B$, is $B \sim \sqrt{D/s}$. This is the width of the sigmoidal shape of the traveling wavefront, where behind and front of the wavefront has $\phi = 1$ and $\phi$ = 0, respectively (\Cref{fig2}) \cite{murray2002mathematical}. The FK equation belongs to a broader class of PDEs that arise as continuum limits of generalized random-walk models in mathematical biology \cite{durrett1994,durrett2014,coxdurrett2016,cheng2026}

This framework has been successfully applied to experimental microbial growth settings \cite{gandhi2016range,datta2013range,baym2016spatiotemporal}. The generalizations of FK incorporate other biologically relevant mechanisms, such as quorum sensing, chemotaxis, and resource depletion, among others. Quorum sensing introduces density-dependent feedbacks, where cells regulate their behavior based on local population density \cite{maslovskaya5388371allen,levin2012mathematical}. Chemotactic models add directed movement to chemical gradients, modifying front shapes and increasing invasion speed \cite{murray2002mathematical,greulich2012mutational,hermsen2012rapidity}. Other extensions include resource-coupled reaction-diffusion systems, spatially explicit competition over shared nutrients or territory, and nonlinear interactions such as public goods dynamics, mechanical pushing between cells, or long-range interactions in certain bacteria \cite{lambert2014bacteria,pigolotti2010coexistence,kun2013resource,Gorgi2026GeometricOrdering}.  

Although the assumption of independent random movement in the FKPP framework is appropriate for in vitro growth assays and microfluid experimental setups, it is less justified in crowded environments such as multicellular tissue structures \cite{komarova2006spatial}. A central example is the somatic evolution in epithelial tissues, which underlies many models of cancer dynamics \cite{manem2015modeling,komarova2006spatial}. In epithelial structures, homeostatic mechanisms enforce space-filling constraints, ensuring that there are no voids and leading to a densely packed, mechanically constrained environment. Cell movement in such tissues is tightly regulated and is typically coupled to local proliferation or apoptosis events among neighboring cells. Selection dynamics in these ``crowded environments" are more suitably described by a DB or BD Moran process.

Another feature of the spatial Moran BD and DB processes is the fact that the dynamics can incorporate two fitness components, that is, the reproduction rate or birth rate (fecundity) and death rate (viability or survival) \cite{kaveh2014duality}. The FKPP and diffusion-reaction frameworks do not make this distinction: birth rate and death rate enter only through the combined selection coefficient.

A prominent example where this separation is essential is the evolution of antimicrobial and drug resistance, where antibacterial agents act through two distinct physiological mechanisms: bacteriostatic or cytostatic effects primarily suppress proliferation, whereas bactericidal or cytotoxic effects increase mortality \cite{Greulich2012,Hermsen2012,Greulich2015,Meredith2015,Coates2018,Baquero2021}. A drug that suppresses growth and a drug that increases mortality can act through entirely different physiological routes while producing the same net fitness difference, a distinction that FKPP-type reaction--diffusion models, which collapse both routes into a single growth parameter \cite{murray2002mathematical,ElHachem2019}, cannot capture. The two-fitness component scenarios recurs broadly in evolutionary biology, including cancer cell fitness, life-history trade-offs between reproduction and survival, and ecological competition under predation versus resource limitation.

Spatial Moran dynamics have been extensively studied for birth-death (BD) and death-birth (DB) update rules \cite{hindersin2015most,zukewich2013consolidating,manem2015modeling,manem2014spatial}; however, mapping such stochastic discrete models to their corresponding continuum space-time partial differential equation (PDE) formulations has received relatively little attention in the literature, with the notable exception of recent work by Houchmandzadeh and Vallade \cite{houchmandzadeh2017fisher}.

Recent work in evolutionary graph theory has examined fixation in graph-structured meta-populations and island networks \cite{kaveh2014duality,Yagoobi2021,Yagoobi2023}. Yagoobi and Traulsen \cite{Yagoobi2021} showed that migration probability and patch-size heterogeneity determine whether such structures amplify or suppress selection, and this was extended in \cite{Yagoobi2023}, which categorized update mechanisms and showed how fixation depends on the order of birth and death events and whether selection acts at birth or death.

In the current work, we present the deterministic limit of the spatial Moran BD or DB process in large island lattices and derive the corresponding partial differential equation in the weak selection limit. We use a two-fitness model, where the mutant birth rate (fecundity) $r=1+s$ and the death rate (viability) $d=1+q$ are both variable, with $s$ and $q$ as the corresponding birth- and death-selection coefficients. We show that the partial differential equations for BD and DB differ from FK, and only in the case $s=0$ and $q\ll1$ does DB reduce to FK. Unlike the FK formalism, the overall selection coefficient, $f=s-q$, is not the parameter that drives the dynamics: the shape of the wavefront and the speed of the wave differ for different values of $s$ and $q$ even when $f$ is held constant.

We numerically compute the Fisher wave speeds and derive an algebraic formula for the speed in the weak selection limit as an implicit function of time. We observe that the resulting selective wave speeds differ substantially in BD, DB, and FK dynamics. When fecundity drives the dynamics, the selective waves decelerate for the BD process; whereas, in the DB process, the wave propagates with a higher, constant speed. In contrast, when viability drives the process, the DB wave accelerates while the BD and FK waves maintain comparable constant speeds. An approximate duality is observed: switching to the case $s=0$, the behavior of DB qualitatively matches that of BD for $q=0$.

We generalize our framework to heterogeneous media and derive a partial differential equation that captures the selection dynamics in heterogeneous continuum media, obtained as the continuum limit of evolutionary dynamics on heterogeneous graphs. Within this framework, we identify a class of media that we term isothermal media, which corresponds to the continuum analog of discrete isothermal graphs. We further demonstrate that isothermal media are mathematically connected through gauge transformations \cite{jackson1998classical} in the functional space of $\mu(x)$ and $\boldsymbol{\alpha}(x)$, which represent spatially varying migration rates and drift fields, respectively.

Our new formalism establishes a framework to study spatial evolutionary processes in compact settings such as epithelial tissue structures, and allows for the incorporation of more complex mechanisms, such as cell-cell interactions, cooperative behavior, environmental interactions, and more complex forms of motility. Some of these extensions are already underway and are the subject of future publications.

\FloatBarrier

\section{Lattice Moran dynamics in continuous space-time limit}

We use an island network model in which a mutant population (genotype or strategy A) competes with a resident population (type B) across a set of connected demes or islands. The population structure is encoded by a graph $G$, represented by the adjacency matrix $m_{ij}$, where $m_{ij} = 1$ if nodes $i$ and $j$ are connected and $m_{ij} = 0$ otherwise. Each island maintains a constant population size $N$. The number of mutants (residents) on island $i$ is $n_i$ ($N - n_i$) respectively. We focus on 1d and 2d lattice graphs (cycle and square lattice) with $K$ islands. The reproduction rate of mutants (residents) is $r_{\rm A} (r_{\rm B})$, respectively. Similarly, the death rate of mutants (residents) is $d_{\rm A}$ ($d_{\rm B}$), respectively (\Cref{fig2}). For dynamics, we consider: $1)$ Moran death-birth update rule (DB); and $2)$ Moran birth-death update rule (BD). We start with a stochastic formulation of these update rules and then present the large-island limit, where the stochasticity is suppressed (` the thermodynamic limit' in statistical physics term). Without loss of generality, we set $r_{\rm A}=r, d_{\rm A}=d, r_{\rm B}=d_{\rm B}=1$. \\

\nd {\bf Death-birth updating.} For the DB process, at each time step, an individual is selected for death with probability proportional to its death rate, $d_{\rm A,B}$. Then a neighboring individual is selected, with probability proportional to its reproduction rate (birth rate) $r_{\rm A,B}$, to reproduce and fill the vacancy on the focal island. This defines a constant-population Markov process in which the number of mutants on a given island increases or decreases by one per update step. The transition probabilities $P^+_{\rm DB}(i)$ $(P^-_{\rm DB}(i) )$ denote the probability that the number of mutants on island $i$ increases (decreases) by one, respectively \cite{nowak2006evolutionary, traulsen2009stochastic, kaveh2014duality}. By definition and due to the constant population constraint, death, birth, and migration events are intrinsically coupled. This modeling structure reflects highly regulated multicellular structures in which this coupling occurs naturally, for example, in epithelial tissues, where cell death initiates a local signaling cascade that drives neighboring cells to proliferate and fill the gap.

Using the standard steps in the literature (see \cite{gardiner2004hanDBook}) we can write the equation for the average mutant frequency, $\phi_i \equiv \langle n_i/N \rangle$ in terms of transition of probabilities: 

\begin{align}
\phi_{i}(t+1)-\phi_{i}(t) &= \frac{1}{N} \left( \langle P^{+}_{\rm DB}({\bf n}/N)\rangle - \langle P^{-}_{\rm DB}({\bf n}/N) \rangle \right)
\nonumber\\
&\approx \frac{1}{N} \left( P^{+}_{\rm DB}({\boldsymbol \phi}) - P^{-}_{\rm DB}({\boldsymbol \phi}) \right),
\label{eq_ave_DB}
\end{align}

\nd where $\langle \cdot \rangle$ denotes the expected value. In the second line we used large-$N$ approximation, $\langle P(a) \rangle \approx P(\langle a \rangle )$. (See the SI for details.)

 For brevity, we focus on the 1d large-cycle with periodic boundary conditions. For a 1d cycle, the adjacency matrix is $m_{ij} = \left( \delta_{i,j+1} + \delta_{i,j-1}\right)$, where $\delta_{ij}$ is the Kronecker delta-function. The \cref{eq_ave_DB} becomes,
 
 \begin{align}
\phi_{i}(t+1)-\phi_{i}(t) &=  \frac{{\rm r}(1-\phi_i(t))}{N(1+({\rm d}-1)\overline{\phi}(t))}\Big(\frac{\phi_{i+1}(t)+\phi_{i-1}(t)}{2+({\rm d}-1)(\phi_{i+1}(t)+\phi_{i-1}(t))}\Big) \nonumber \\ &- \frac{{\rm d}\phi_i(t)}{N(1+({\rm d}-1)\overline{\phi}(t))}\Big(\frac{2-(\phi_{i+1}(t)+\phi_{i-1}(t))}{2+({\rm d}-1)(\phi_{i+1}(t)+\phi_{i-1}(t))}\Big).
\label{DB_rd}
\end{align}

To obtain the transition probabilities in \cref{DB_rd}, we decompose each update event into the corresponding sequence of elementary conditional steps defined by the update rule. Specifically, for the BD and DB dynamics, the probabilities $(P_{\rm BD})$ and $(P_{\rm DB})$ are constructed by combining the conditional probabilities for the constituent steps of the event sequence (selection of the death event, selection of the reproducing event, and migration/replacement), together with the appropriate normalization factors. The complete factor-by-factor derivation, including the source-to-target forms $(P_{\rm BD}(j\to i))$ and $(P_{\rm DB}(j\to i))$, is provided in the Supplementary Information.

The underlying Moran process is a discrete-time Markov chain, with one BD or DB replacement event per update step. The PDE is obtained from a macroscopic scaling limit of this discrete process: we introduce a rescaled time increment \(\Delta t = 1/(KN)\) and lattice spacing \(\Delta x = 1/K\), and then take the large-system limit \(K,N \to \infty\), so that \(\Delta t,\Delta x \to 0\). Thus, \(\partial_t \phi\) refers to the rescaled continuum time variable rather than a continuous microscopic time. We map the spatial index of the lattice, $i$, to the 1d or 2d continuum space ${\bf x}$, and thus $\phi_i \to \phi({\bf x})$. We introduce two selection coefficients: ${\rm s}=r-1$ for birth and ${\rm q}=d-1$ for death. In the weak-selection regime ($s, q \ll 1$), \cref{DB_rd} simplifies to:


\begin{align}
 \frac{\partial \phi({\bf x},t)}{\partial t}  &= D\Big\{\big(1+{\rm s}- {\rm q}\overline{\phi}\big) - \big(2{\rm s} - {\rm q}\big)\phi\Big\}\nabla^2\phi({\bf x},t) \nonumber\\
&+ \big({\rm s} - {\rm q}\big) \phi ( 1 - \phi ) + \mathcal{O}\Big({\rm s}^2 ,{\rm q}^2, \nabla^4 \Big),
~~~~~~~~~~\textrm{(DB)}
\label{DB_weak}
\end{align}

\nd where $\nabla^2\phi \approx \Delta_i^2 \phi/(\Delta x)^2$ and $\partial \phi/\partial t \approx \Delta \phi_i/(\Delta t)$ ($\Delta_i^2$ and $\Delta_i$ are the second and first discrete derivatives.) The diffusion constant $D$ is defined as $D = (1/2)(\Delta x)^2 = 1/(2K^2)$. $\overline{\phi}= \int_{x} \phi(x,t) dx$  is the mean fitness across the population. We use $\nabla$ to emphasize that the above derivation is also true in two dimensions (square lattice) as well (see SI). Higher-order terms \(\mathcal{O}({\rm s}^2 ,{\rm q}^2, \nabla^4)\) include quadratic corrections in the selection coefficients, as well as higher spatial derivatives neglected under the weak selection limit. \\

\nd {\bf Birth–death updating.} In the BD scheme, each time step begins with the selection of an individual to reproduce, with probability proportional to its fitness (a global event). The resulting offspring then replaces an individual on a neighboring island, selected with probability proportional to the neighbor’s death rate. As with the DB update, this defines a Markov process in which the number of mutants on a given island changes by one per time step. We denote the corresponding transition probabilities for the BD process with $P^+_{\rm BD}(i)$ $(P^-_{\rm BD}(i) )$.

Similarly to the DB case, the discrete equation for the average frequency $\langle \boldsymbol{\phi} \rangle$ in the 1d cycle graph is given by

\begin{align}
&\phi_{i}(t+1)-\phi_{i}(t) =  \nonumber\\
&\frac{{\rm r}(1-\phi_i(t))}{N(1+({\rm r}-1)\overline{\phi}(t))}\Big( \frac{\phi_{i-1}(t)}{2+({\rm d}-1)\phi_{i}(t)+\phi_{i-2}(t)} 
+ \frac{\phi_{i+1}(t)}{2+({\rm d}-1)\phi_{i}(t)+\phi_{i+2}(t)} \Big),
\nonumber \\
&- \frac{{\rm d} \phi_i(t))}{N(1+({\rm r}-1)\overline{\phi}(t))}\Big( \frac{1-\phi_{i-1}(t)}{2+({\rm d}-1)\phi_{i}(t)+\phi_{i-2}(t)} + \frac{1-\phi_{i+1}(t)}{2+({\rm d}-1)\phi_{i}(t)+\phi_{i+2}(t)} \Big) \nonumber\\
\label{BD_rd}
\end{align}

The weak-selection continuum limit is given by the partial differential equation, 

\begin{align}
 \frac{\partial \phi({\bf x},t)}{\partial t}  &=  D\Big\{\big(1+{\rm s} - {\rm s}\big(\phi+\overline{\phi}\big)\Big\}\nabla^2\phi({\bf x},t) + 2{\rm q}D \big(\nabla \phi\big)^2 \nonumber\\
&+ \big({\rm s} - {\rm q}\big) \phi ( 1 - \phi ) + \mathcal{O}\Big({\rm s}^2 ,{\rm q}^2, \nabla^4\Big).
~~~~~~~~~~\textrm{(BD)}
\label{BD_weak}
\end{align}

In this case, we now have an additional KPZ-type term, $(\nabla \phi)^2$, which can affect pattern formation of mutant colonies. However, this term does not change the selective wave speed for the BD process. 

As evident from the expressions above, the dynamics under both DB and BD updating include a global term $\overline{\phi}$, the mean mutant frequency in all islands. This term introduces a long-range interaction similar to a mean field between local mutant frequencies $\phi_i$, effectively coupling the dynamics across space. This structure allows us to study the emergence of selective wave patterns and compute wavefront propagation speeds. In particular, we can compare the resulting wave behavior with that predicted by the classical FKPP equation.\\


\section{Traveling-wave solutions and Fisher wave speeds for BD and DB processes}

\begin{figure}
\begin{center}
\includegraphics[width=1 \textwidth]{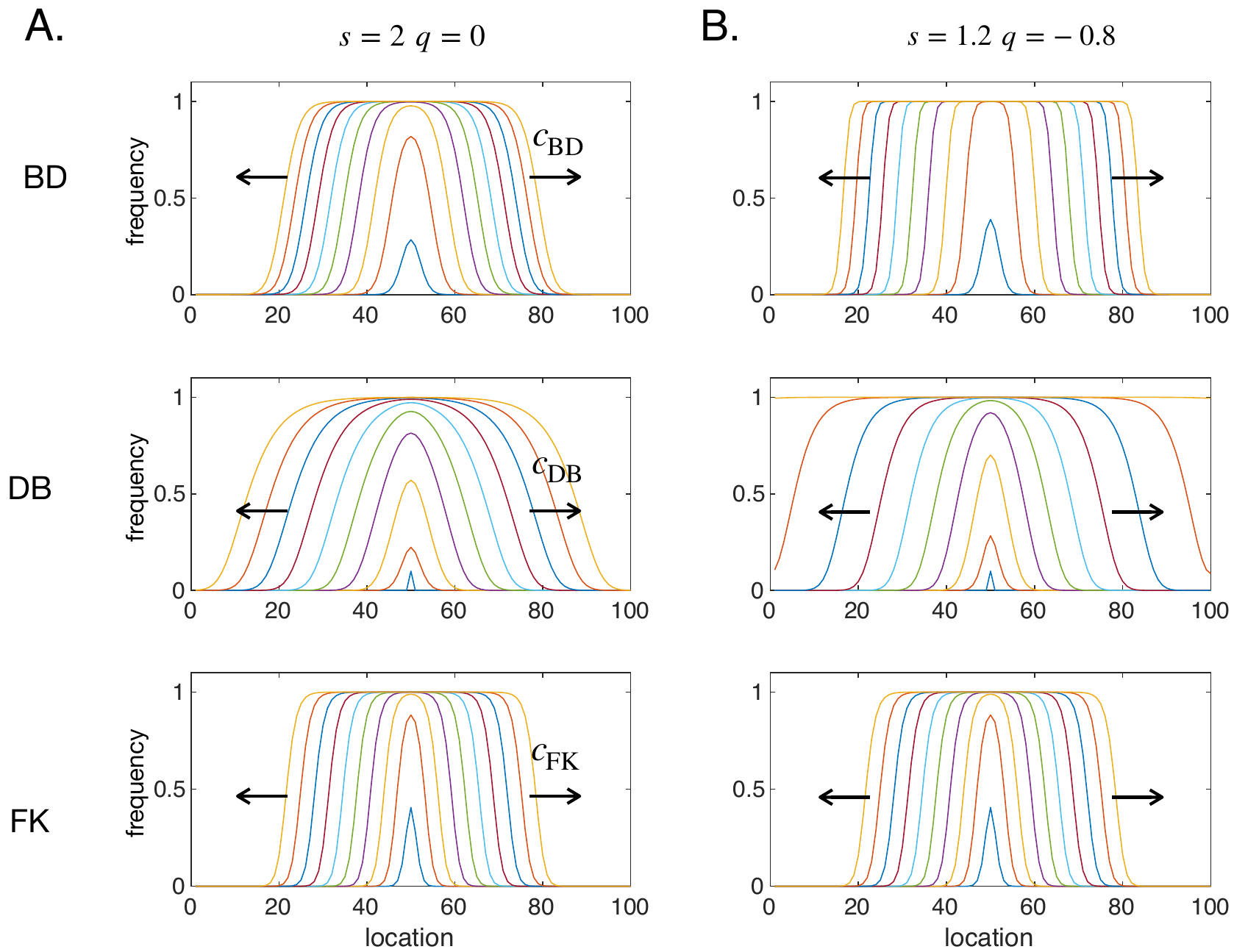}
\end{center}
    \caption{
        \textbf{Wavefront propagation in BD, DB, and FK dynamics with equivalent net selection coefficient.} 
        Panels show snapshots of the mutant frequency \(\phi(x,t)\) over time for three models: birth-death (BD), death-birth (DB), and Fisher-KPP (FK), on a 1D lattice of \(K = 100\) islands with a single initial mutant at the center.
        \textbf{(A)} For selection parameters \(s = 2, q = 0\), corresponding to pure birth advantage. 
        \textbf{(B)} For selection parameters \(s = 1.2, q = -0.8\), where the net selective advantage \(s - q = 2\) remains constant. 
        Each row compares wavefront speeds \(c_{\mathrm{BD}}, c_{\mathrm{DB}}, c_{\mathrm{FK}}\), showing that DB spreads faster than BD and FK in both scenarios, and that wavefront curvature/width differs across update rules despite identical net selection coefficient.
    }
\label{fig3}
\end{figure}

We now investigate the traveling wave solutions for the two update rules \cref{BD_rd,DB_rd} and / or their continuous-media versions, \cref{DB_weak,BD_weak}. Numerical solutions show a traveling wave pattern similar to the FK dynamics. However, the speed, shape, and width of the wavefront are not function of the overall fitness coefficient, that is, ${\rm r} - {\rm d} = {\rm s} - {\rm q}$. In \Cref{fig3}, the wavefront is plotted for various time points for a 1d lattice graph of $K=100$ and $N=100$. The results for two sets of parameters $(s=2, q=0)$ and $(s=1.2, q=-0.8)$ are shown, where both cases have the same overall fitness coefficient $s-q$. We assume a single mutant started on the middle island, $i=50$ at $t=0$ and created two traveling wavefronts to the left and right. DB and BD dynamics differ in speed and wavefront shape between panel A ($s=2,q=0$) and panel B ($s=1.2, q=-0.8$). More interestingly, DB propagates faster than BD and FK in both cases. 


An advantage of partial differential equations \cref{BD_weak,DB_weak} is that we can use existing literature on FK to obtain approximate analytical results for the selective wave speeds of the BD and DB dynamics. A similar analytical estimate for the Fisher wave speed in an agent-based model, roughly corresponding to our DB model with ${\rm d}=1$, has been studied in \cite{houchmandzadeh2017fisher}. Following \cite{houchmandzadeh2017fisher, murray2002mathematical}, we can derive similar formulas for the wave speed in weak selection (see SI for details):

\begin{align}
c_{\rm BD}(\overline{\phi}) &\approx 2\sqrt{D\Big(s - q\Big)\Big((1 + s) - s \overline{\phi}\Big)}\nonumber\\
c_{\rm DB}(\overline{\phi}) &\approx 2\sqrt{D\Big(s - q\Big)\Big((1 + s ) - q\overline{\phi} \Big)}.\nonumber\\
\label{speed_equation}
\end{align}

Notice that our results depend on the mean mutant frequency $\overline{\phi}$, which is indicative of an accelerating (or decelerating) wave. These results should be compared with those of the FK results $c_{\rm FK} = 2\sqrt{Ds}$ (see \cite{murray2002mathematical}). 

\begin{figure}[h]
\begin{center}
\includegraphics[width=1\textwidth]{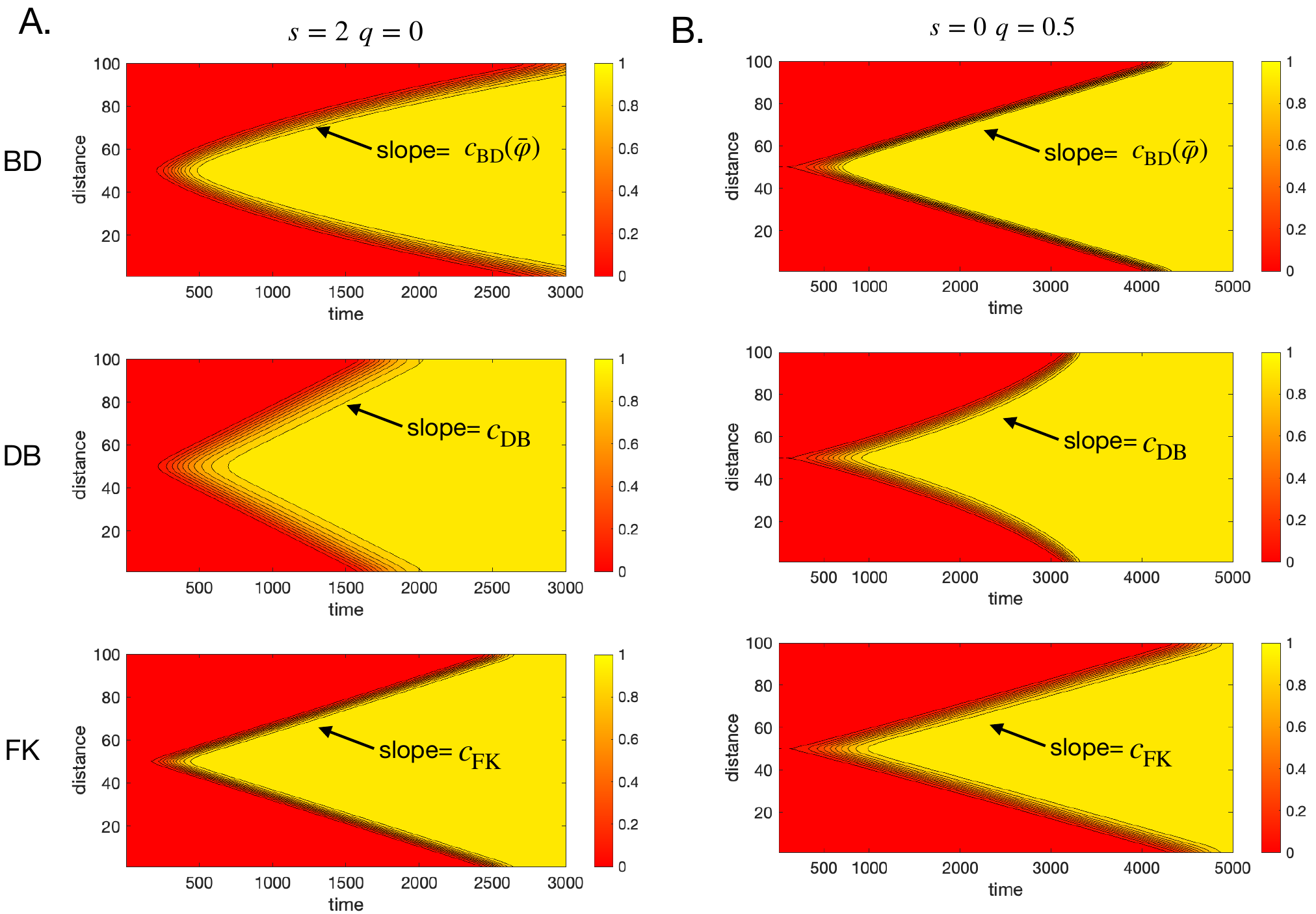}
\end{center}
\caption{
    \textbf{Spatiotemporal dynamics of selective wavefronts in 1D for BD, DB, and FK models.} 
    Each panel shows the spatiotemporal heatmap of mutant frequency \(\phi(x,t)\) on a 1D lattice with \(K=100\) islands, initiated by a single central mutant. Contours correspond to constant frequency levels.
    \textbf{(A)} For \(s = 2, q = 0\), representing pure birth selection. The wavefront in the BD model exhibits deceleration (nonlinear slope \(c_{\mathrm{BD}}(\bar{\phi})\)), while DB and FK show more consistent wave speeds, with DB propagating fastest.
    \textbf{(B)} For \(s = 0, q = 0.5\), representing pure death selection. Here, BD wave speed is more stable, while DB displays curvature due to frequency-dependent acceleration. In both cases, the DB dynamics yield the highest wave speed compared to BD and FK, despite equal net selection \(s - q\).
}
\label{fig4}
\end{figure}

\begin{figure}
\centering
\includegraphics[width=0.9\textwidth]{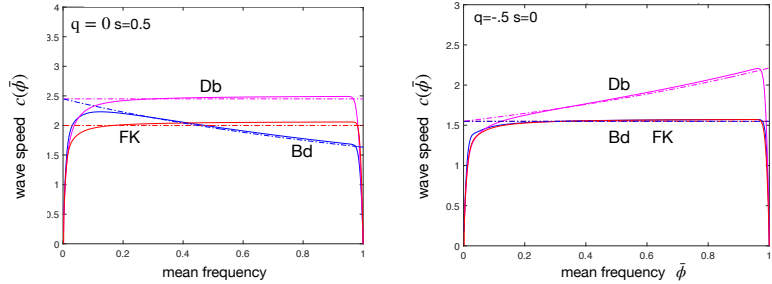}
\caption{
    \textbf{Frequency-dependent wave speeds in birth-death, death-birth, and Fisher-KPP dynamics.} 
    Wave propagation speeds \(c(\bar{\phi})\) are plotted against the mean mutant frequency \(\bar{\phi}\) for three models: birth-death (BD), death-birth (DB), and Fisher-Kolmogorov (FK). 
    \textbf{Left:} When selection arises solely through increased birth rate (\(s = 0.5, q = 0\)), DB yields the fastest wave and exhibits weak acceleration, while BD displays mild deceleration. 
    \textbf{Right:} When selection arises via decreased death rate (\(q = -0.3, s = 0\)), the DB wavefront accelerates strongly with frequency, in contrast to the nearly constant speeds of BD and FK. 
}
\label{fig5}
\end{figure}

\cref{speed_equation} shows that the propagation speed is not determined solely by the net coefficient $s-q$. While $s-q$ sets the FK-like baseline scale, the BD/DB update rule introduces a multiplicative prefactor that depends separately on $s$ and $q$, and therefore on the mechanism by which selection acts (birth- versus death-driven). As a result, parameter sets with the same (s-q) can produce different wave speeds and profile shapes. This update rule-dependent correction accounts for the deviations from FK behavior observed in \Cref{fig4,fig5} (including acceleration/deceleration), and the full derivation is given in the SI (Eq. S.28).

\FloatBarrier

\FloatBarrier

A natural complement to the selective wave speed is the fixation time,
\(T_{\text{fix}}\), defined here as the characteristic time required for the
mutant wave to traverse the system. This quantity can be read directly from
the space--time cone: different cone slopes correspond to different wave
speeds, and therefore to different fixation times. Formally, for a system of
spatial extent \(L\), we define
\begin{equation}
    T_{\text{fix}} \;=\; \frac{L}{\bar{c}},
    \label{eq:Tfix}
\end{equation}
where \(\bar{c}\) is the time-averaged wave speed.

Because the BD, DB, and FK dynamics generate different space--time cones,
their corresponding fixation times are also update-rule dependent. In the BD
and DB processes, \(\bar{c}\) can additionally depend on the mean mutant
frequency through \cref{speed_equation}, so \(T_{\text{fix}}\) depends on
the individual values of \(s\) and \(q\), not merely on their difference
\(s-q\). Thus, even parameter sets with the same net selective advantage can
produce different fixation times depending on whether selection acts through
birth, death, or the update rule itself. Examples of strong-selection regimes,
including cases with large \(s\) and small \(q\), are provided in the
SI (Figure S2) . These examples further illustrate that the separation
between \(s\) and \(q\) becomes increasingly important outside the weak-selection
limit, where one-parameter approximations based only on \(s-q\) can fail.


The space-time cone representation of the traveling wave solutions in BD, DB and FK scenarios is presented in \Cref{fig4}. These are the solutions for the discrete \cref{DB_rd,BD_rd} and the discretized version of FK on a lattice of 100 islands ($K=100$) and each island capacity is $N=100$. In panel A, the reproductive fitness is different between two types, that is, $s \neq 0, q=0$, and the curvature of the BD cone signals a frequency dependence or an implicit time dependence of the wave speed. (The slope of the space-time cone is the wave speed.) In this case, DB seems to have a constant speed that is larger than that of FK. This can be seen in analytical solutions, \cref{speed_equation}, where $c_{\rm DB}(q=0) = 2\sqrt{Ds(1+s)} > 2\sqrt{sD} = c_{\rm FK}$. Interestingly, in the case where the death selection coefficient drives the dynamics, $q \neq 0, s=0$, the frequency dependence of the speed is on DB, and the space-time cone is now convexly curved. However, in our observations for both sets of the parameters chosen, DB has the fastest selective wave compared to those of FK and BD. This is indicative of a qualitative duality between the BD and DB processes.

In  \Cref{fig5}, the frequency dependence of the BD and DB wave speeds in the above two scenarios, $(s\neq 0, q=0)$ and $(s=0, q\neq 0)$ as a function of the mean frequency, $\overline{\phi}$ is depicted. As can be seen, on the figure,  the analytical results, \cref{speed_equation} matches well with the numerical results of \cref{DB_rd,BD_rd} in this weak/medium selection regime. We also plotted the average Fisher wave speeds for different $s$ and $q$ in three models/update rules, BD, DB and FK. For the results (see Figure S1, panels B and C, in the SI).

\section{Continuous-media analog of heterogeneous graphs: isothermal gauge transform}

Biological populations are rarely spatially uniform. In bacterial communities and biofilms, cells experience nonuniform density, crowding, nutrient availability, and connectivity; in epithelial tissues, somatic evolution unfolds within spatially constrained, heterogeneous neighborhood structures. We extend the lattice-based BD/DB derivation above to this setting by allowing the migration and replacement mechanisms introduced earlier to vary across space. Evolutionary graph theory offers a natural representation for such systems, in which individuals occupy nodes and replacement or dispersal events occur along edges \cite{lieberman2005evolutionary,nowak2006evolutionary}; heterogeneous graphs can thus be viewed as discrete approximations of spatially structured biological media whose local connectivity, motility, or replacement rates vary across position.

Within evolutionary graph theory, prior work on heterogeneous structures has focused mainly on fixation probability: a general heterogeneous graphs can amplify or suppress selection depending on topology and update rule. This is distinct from the heterogeneous reaction-diffusion literature, where spatially varying diffusivity or growth rate is well studied in relation to front speed and front shape. Our continuum limit model starts by coarse graining a heterogeneous graph into a continuum media and construct a continuum space-time dynamics as before, now new environmental fields introduced into the PDE to represent the heterogeneity in the continuous medium.

We generalize the uniform model discussed above by including variable motility and bias in the migration matrix. In a uniform 1d lattice (cycle), we define motility, $\mu$, and bias, $a$, using a reparameterization of the elements of the migration matrix: $m_{i,i\pm1} = (\mu \pm a) /2 $ and, $m_{ii}=1-\mu$. $\mu$ is the probability that an offspring residing in island $i$ moves out and $1-\mu$ is the probability that it stays in. The bias $a$, is the migration bias for moving toward the right. In a uniform 2d square lattice, bias is a two-dimensional vector ${\bf a}$. Its first (second) component is half the difference between the migration probability towards the right/left (up/down), respectively. The bias may be due to irregularities in the tissue structure or the result of a chemotaxis gradient. We will focus on the BD process for now and leave a more thorough investigation of heterogeneous media to future work. 

For a 1d heterogeneous or weighted cycle, we similarly introduce the parameters $\mu_i, \alpha^{R}_i, \alpha^{L}_i$ and write the migration matrix elements as: $m_{i,i+1} = \frac{1}{2} \big( \mu_i + \alpha^{\rm R}_i \big),  m_{i,i-1} = \frac{1}{2} \big( \mu_i + \alpha^{\rm L}_i \big)$ and $m_{ii} = 1- \mu_i$. $\mu_i$ is the location-dependent motility on the island $i$. The parameters $\alpha^{\rm R}_i (\alpha^{\rm L}_i)$ are the offset probabilities that an offspring of the island $i$ migrates to the right (left), respectively. Since $m_{i,i+1}+m_{i,i-1}+m_{i,i} = 1$ we have $\alpha^{\rm L}_i +\alpha^{\rm R}_i =0$.  The bias is $a_i = \alpha^{\rm R}_i-\alpha^{\rm L}_i$.

In 2d we use two components for the lattice index: ${\bf i} = (i,j)$ and similarly, the migration matrix $m_{ij,kl \pm 1}$ is written in terms of five parameters: $\mu_{\bf i}, \alpha^{\rm R}_{\bf i},\alpha^{\rm L}_{\bf i},\alpha^{\rm U}_{\bf i},\alpha^{\rm D}_{\bf i}$ where $\alpha^{\rm U} + \alpha^{\rm D} + \alpha^{\rm R} + \alpha^{\rm L} =0$: $m_{ij,i+1j} = \frac{1}{4}\big(\mu_{\bf i} + \alpha^{\rm R}_{\bf i}), m_{ij,i-1j} = \frac{1}{4}\big(\mu_{\bf i} + \alpha^{\rm L}_{\bf i}),m_{ij,ij+1} = \frac{1}{4}\big(\mu_{\bf i} + \alpha^{\rm U}_{\bf i}),m_{ij,ij-1} = \frac{1}{4}\big(\mu_{\bf i} + \alpha^{\rm D}_{\bf i})$.

\begin{figure}[h]
\includegraphics[width=0.9\textwidth]{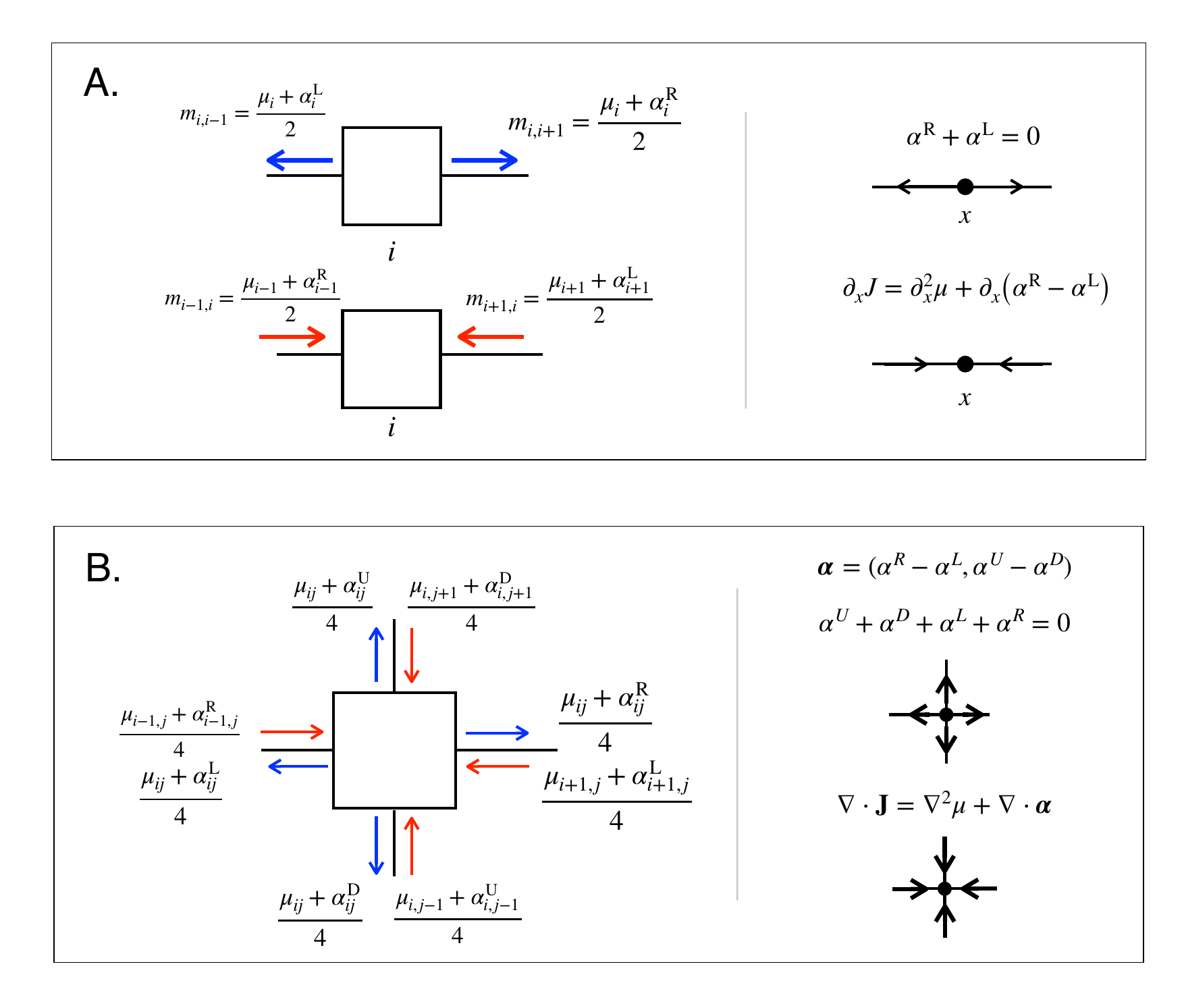}
    \caption{
        \textbf{Migration matrices and isothermal constraints in heterogeneous spatial media.}
        \textbf{(A)} One-dimensional case: Schematic of local migration probabilities in a heterogeneous cycle, where motility \(\mu_i\) and directional biases \(\alpha_i^R, \alpha_i^L\) determine transition rates between neighboring sites. The isothermal constraint requires that \(\alpha_i^R + \alpha_i^L = 0\), ensuring conservation of local migration probability (node “temperature”). The resulting conservation law is expressed as \(\partial_x J = \partial_x^2 \mu + \partial_x(\alpha^R - \alpha^L)\).
        \textbf{(B)} Two-dimensional case: Migration fluxes in a square lattice, with directional components \(\alpha_{ij}^U, \alpha_{ij}^D, \alpha_{ij}^L, \alpha_{ij}^R\) and local motility \(\mu_{ij}\). The isothermal condition generalizes to \(\alpha^U + \alpha^D + \alpha^L + \alpha^R = 0\), and the corresponding conservation law becomes \(\nabla \cdot \mathbf{J} = \nabla^2 \mu + \nabla \cdot \boldsymbol{\alpha}\). These relations ensure that the total incoming migration to a site remains balanced, analogous to the isothermal theorem for graphs.
    }
\label{fig6}
\end{figure}

Using the above parameterizations, motility $\mu_i$ and bias ${\boldsymbol \alpha}_i$, we can map the discrete heterogeneous graph to a continuum limit of a heterogeneous graph, which is now identified by the scalar and vector fields $\mu({\bf x})$ and ${\boldsymbol \alpha}({\bf x})$ (\Cref{fig6}).

Following a similar derivation as in the uniform case, we can derive a general PDE that governs the selection dynamics in heterogeneous media for BD dynamics. The 1d version of the results is as follows:

\begin{align}
\frac{\partial\phi(x,t)}{\partial t}&=
\big[(1+s-2s\phi(x,t))-q(\overline{\phi}(x,t)-\phi(x,t)\big]\Big(D\mu(x)\partial_{x}^2\phi(x,t)\nonumber\\ 
&+\big(v\alpha(x)+2D(\partial_x \mu(x)-\alpha(x) \partial_{x}\alpha(x))\big)\partial_{x}\phi(x,t)\Big) \nonumber\\
&+(s-q)\phi(x,t)(1-\phi(x,t))-2Ds\alpha(x)^2(\partial_{x}\phi(x,t))^2,
\label{equation8}
\end{align}

\nd where $\partial_x$ denotes the spatial derivative. As demonstrated in \cref{equation8}, environmental heterogeneity influences invasion dynamics through two distinct mechanisms: (i) a spatially varying effective diffusivity via the factor $\mu(x)$ multiplying the second-derivative term, and (ii) a drift contribution involving $\alpha(x)$ together with gradients of $\mu(x)$ and $\alpha(x)$. As a result, selective invasion fronts in heterogeneous media are not expected to be governed solely by the homogeneous-wave picture: local increases (decreases) in effective motility can accelerate (decelerate) front propagation, while directional bias can skew or deform the front and induce asymmetric propagation. Thus, even at fixed $(s,q)$, the expansion velocity and front morphology generally become environment-dependent functionals of $\mu(x)$ and $\alpha(x)$.

A systematic analysis of traveling-wave solutions and front speeds in specific classes of heterogeneous media is an important next step, but lies beyond the scope of the present work and will be treated in a future publication.

{\bf Motility as a gauge transform.} A crucial observation in evolutionary dynamics in spatial graph structures is the isothermal theorem. It states that for a category of structures known as isothermal graphs, the fixation probability of a randomly placed mutant is independent of the spatial structure. The isothermal graph is a weight graph $m_{ij}$, where the migration matrix between node $i$ and $j$ has the incoming and outgoing weights sum up to unity. This includes all undirected regular graphs and lattice graphs. 

There is significant literature on the probability of fixation in graph structures and extensions of the isothermal theorem to heterogeneous graphs \cite{adlam2014universality} and heterogeneous environments \cite{kaveh2019environmental}, and variable birth and death models \cite{kaveh2020moran}. The total incoming weight or migration probability into each node of an evolutionary graph is called the `temperature of that node. An isothermal graph is a graph for which the temperature of each node is equal to unity. For the birth-death (BD) processes, the sum of outgoing migration in each node is also equal to unity. Thus, the migration matrix for an isothermal graph is a doubly stochastic matrix. The isothermal theorem \cite{baym2016spatiotemporal,maruyama2} states that the fixation probability of a randomly placed mutant on any isothermal graph is the same and equal to the well-known Moran fixation probability in unstructured populations. 

We call the continuum limit version of an isothermal graph an `isothermal medium'.  It is straightforward to check that in isothermal media the isothermality condition leads to a conserved current written as:

\begin{align}
\nabla \cdot \big({\boldsymbol \alpha}({\bf x}) + \nabla \mu({\bf x})\big) = 0
\end{align}

In other words, the vector ${\boldsymbol \alpha}({\bf x})$ is related to other isothermal graphs up to a gauge transform ${\boldsymbol \alpha} \to {\boldsymbol \alpha} + \nabla \mu$. However, not every two isothermal media are related by such gauge transform.

\section*{Discussion}

We presented a continuum framework for spatial Moran processes, deriving partial differential equations that govern selection dynamics under both birth–death (BD) and death–birth (DB) update schemes.
The results intorduces a new category of diffusion-reaction PDEs with nonlinear and nonlocal diffusion terms. Our approach is a complementary perspective on how continuum PDE models can emerge from discrete stochastic spatial dynamics in mathematical biology \cite{Codling2008,Plank2025,Keener2021}. Our results show that, although these two processes appear superficially similar, they generate markedly different spatiotemporal dynamics when extended to continuous space, especially in the presence of distinct birth and death selection coefficients. Following \cite{kaveh2014duality} we use a two-component fitness model, where fecundity (birth rate) and survival (death rate) are considered independent variables for each genotype. Comparisons between the fixation probabilities of a randomly placed mutant in the BD and DB processes have been discussed in the past \cite{kaveh2014duality}.

A key finding in the current work is that the net fitness difference, $f=s-q$, commonly used in well-mixed models, does not fully characterize selective wave propagation in spatial settings. Instead, individual contributions of the birth selection coefficient ($s$), and the death selection coefficient ($q$), determine the wave speed and front morphology in non-trivial ways. In particular, we observe that the DB dynamics consistently yields faster selective wavefronts than the BD or FKPP models, even when the net selection strength is held constant. This result challenges the conventional assumption that the one-component fitness FKPP dynamics adequately captures the general behavior of selection in space and highlights the importance of modeling update mechanisms explicitly. We further observe that when selection is driven by the reproductive fitness coefficient ($s$), the BD dynamics produces decelerating waves, while the DB wave maintains a nearly constant speed. In contrast, when selection is driven by the viability coefficient ($q \neq 0 , s=0$), the situation is reversed: DB waves accelerate, while BD waves remain constant-speed. Only in the latter case are the BD and FK wave speeds identical. (This is not the case for the former.)

The origin of this frequency dependence is worth making explicit, since it is
the central mechanical difference between our model and FKPP. Because each
island maintains a strictly constant population size, birth, death, and
migration are not independent events in the BD or DB update: filling a
vacancy at one site necessarily depends on which individual was selected, at
the population level, to reproduce or die. This constant-population
constraint introduces a global, mean-field-like term, the population-average
mutant frequency $\bar\phi$, into the local transition probabilities (\cref{DB_rd,BD_rd}).
In the continuum limit, this term survives as the $\bar\phi$-dependent
prefactor multiplying the FKPP-like baseline speed in \cref{speed_equation},
rather than being eliminated by the weak-selection expansion. This is distinct
from the frequency dependence already present in the FKPP reaction term
$\phi(1-\phi)$: in our model, the propagation prefactor itself, not
merely the local growth term, changes as the global mutant frequency $\bar\phi$
evolves, which is what produces the accelerating or decelerating wavefronts
in \Cref{fig4,fig5} rather than the constant-speed FKPP front.

Biologically, this constant-population coupling is not a modeling artifact
but a direct consequence of homeostatic regulation. In epithelial tissues,
for instance, cell loss at one location is known to trigger compensatory
proliferation in neighboring cells to maintain tissue integrity, coupling
birth and death events at the level of the whole population rather than
treating them as independent Poisson processes, as in standard reaction-diffusion
derivations. The BD and DB update rules isolate two distinct ways this
coupling can be realized -- selection acting first at a global birth event
versus a global death event -- and our results show that this distinction
alone, independent of the net fitness difference $s-q$, determines whether
the resulting selective wave accelerates, decelerates, or remains constant.

Moreover, we generalize our continuum model to heterogeneous media by developing a continuum analogue of evolutionary graphs with spatially heterogeneous migration and drift. Within this framework, we define ``isothermal media", the continuous counterparts to isothermal graphs, and show that these environments satisfy a conservation law corresponding to a balance of local motility/migration currents. This provides a novel geometric and analytical perspective for studying selection in nonuniform landscapes.

These insights have practical implications for modeling evolutionary dynamics in crowded or spatially constrained systems such as epithelial tissues, microbial colonies, and solid tumors. By moving beyond well-mixed or FK approximations, our approach opens new avenues for incorporating tissue architecture, migration bias, and local environmental variability into models of somatic evolution and adaptation. In particular, this framework can complement recent studies on cancer evolution in brain tumors \cite{waclaw2015spatial,swanson2003virtual,ferreira2002reaction,murray2003spatial}, as well as spatial models of other solid tumors \cite{waclaw2015spatial,noble2022spatial}.

\section*{Data accessibility}
Simulation data and scripts used to generate the results have been deposited in GitHub \cite{MoranDynamicsGitHub}.

\bibliographystyle{elsarticle-num}
\bibliography{citation}   

@book{nowak2006evolutionary,
  title={Evolutionary dynamics},
  author={Nowak, Martin A},
  year={2006},
  publisher={Harvard University Press}
}

@book{broom2014game,
  title={Game-theoretical models in biology},
  author={Broom, Mark and Rycht{\'a}r, Jan},
  year={2014},
  publisher={CRC Press}
}

@article{traulsen2009stochastic,
  title={Stochastic evolutionary game dynamics},
  author={Traulsen, Arne and Hauert, Christoph},
  journal={Reviews of nonlinear dynamics and complexity},
  volume={2},
  pages={25--61},
  year={2009},
  publisher={Wiley-VHC, New York}
}

@book{moran1962,
  title={ The Statistical Processes of Evolutionary Theory},
  author={Moran, P},
  volume={First Edition},
  year={1962},
  publisher={Clarendon, Oxford}
}

@article{maruyama2,
  title={A simple proof that certain quantities are independent of the geographical structure of population},
  author={Maruyama, T},
  journal={Theor. Popul. Biol.},
  volume={5},
  number={2},
  pages={148-54},
  year={1974},
  publisher={Elsevier}
}

@article{lieberman2005evolutionary,
  title={Evolutionary dynamics on graphs},
  author={Lieberman, Erez and Hauert, Christoph and Nowak, Martin A},
  journal={Nature},
  volume={433},
  number={7023},
  pages={312--316},
  year={2005},
  publisher={Nature Publishing Group}
}

@article{adlam2014universality,
  title={Universality of fixation probabilities in randomly structured populations},
  author={Adlam, Ben and Nowak, Martin A},
  journal={Scientific reports},
  volume={4},
  year={2014},
  publisher={Nature Publishing Group}
}

@article{komarova2006spatial,
  title={Spatial stochastic models for cancer initiation and progression},
  author={Komarova, NL},
  journal={Bull. Math. Biol.},
  volume={68},
  number={7},
  pages={1573-1599},
  year={2006}
}

@article{manem2014spatial,
  title={Spatial invasion dynamics on random and unstructured meshes: Implications for heterogeneous tumor populations},
  author={Manem, Venkata SK and Kohandel, M and Komarova, NL and Sivaloganathan, S},
  journal={Journal of theoretical biology},
  volume={349},
  pages={66--73},
  year={2014},
  publisher={Elsevier}
}

@article{hindersin2015most,
  title={Most undirected random graphs are amplifiers of selection for birth-death dynamics, but suppressors of selection for death-birth dynamics},
  author={Hindersin, Laura and Traulsen, Arne},
  journal={PLoS Comput Biol},
  volume={11},
  number={11},
  pages={e1004437},
  year={2015},
  publisher={Public Library of Science}
}

@article{ohtsuki2006simple,
  title={A simple rule for the evolution of cooperation on graphs and social networks},
  author={Ohtsuki, Hisashi and Hauert, Christoph and Lieberman, Erez and Nowak, Martin A},
  journal={Nature},
  volume={441},
  number={7092},
  pages={502--505},
  year={2006},
  publisher={Nature Publishing Group}
}

@article{zukewich2013consolidating,
  title={Consolidating birth-death and death-birth processes in structured populations},
  author={Zukewich, Joshua and Kurella, Venu and Doebeli, Michael and Hauert, Christoph},
  journal={PloS one},
  volume={8},
  number={1},
  pages={e54639},
  year={2013},
  publisher={Public Library of Science}
}

@article{kaveh2014duality,
  title={The duality of spatial death-birth and birth-death processes and limitations of the isothermal theorem},
  author={Kaveh, Kamran and Komarova, Natalia and Kohandel, Mohammad},
  journal={Royal Society Open Science 11/2014; 2(4)},
  year={2014}
}

@article{kun2013resource,
  title={Resource heterogeneity can facilitate cooperation},
  author={Kun, {\'A}d{\'a}m and Dieckmann, Ulf},
  journal={Nature communications},
  volume={4},
  year={2013},
  publisher={Nature Publishing Group}
}

@article{pigolotti2010coexistence,
  title={Coexistence and invasibility in a two-species competition model with habitat-preference},
  author={Pigolotti, Simone and Cencini, Massimo},
  journal={Journal of theoretical biology},
  volume={265},
  number={4},
  pages={609--617},
  year={2010},
  publisher={Elsevier}
}

@article{kaveh2019environmental,
  title={Environmental fitness heterogeneity in the Moran process},
  author={Kaveh, Kamran and McAvoy, Alex and Nowak, Martin A},
  journal={Royal Society open science},
  volume={6},
  number={1},
  pages={181661},
  year={2019},
  publisher={The Royal Society}
}

@article{kaveh2020moran,
  title={The Moran process on 2-chromatic graphs},
  author={Kaveh, Kamran and McAvoy, Alex and Chatterjee, Krishnendu and Nowak, Martin A},
  journal={PLOS Computational Biology},
  volume={16},
  number={11},
  pages={e1008402},
  year={2020},
  publisher={Public Library of Science San Francisco, CA USA}
}

@article{manem2015modeling,
  title={Modeling invasion dynamics with spatial random-fitness due to micro-environment},
  author={Manem, Venkata SK and Kaveh, Kamran and Kohandel, Mohammad and Sivaloganathan, Siv},
  journal={PLoS One},
  volume={10},
  number={10},
  pages={e0140234},
  year={2015},
  publisher={Public Library of Science San Francisco, CA USA}
}

@article{greulich2012mutational,
  title={Mutational pathway determines whether drug gradients accelerate evolution of drug-resistant cells},
  author={Greulich, Philip and Waclaw, Bart{\l}omiej and Allen, Rosalind J},
  journal={Physical Review Letters},
  volume={109},
  number={8},
  pages={088101},
  year={2012},
  publisher={APS}
}

@article{hermsen2012rapidity,
  title={On the rapidity of antibiotic resistance evolution facilitated by a concentration gradient},
  author={Hermsen, Rutger and Deris, J Barrett and Hwa, Terence},
  journal={Proceedings of the National Academy of Sciences},
  volume={109},
  number={27},
  pages={10775--10780},
  year={2012},
  publisher={National Acad Sciences}
}

@article{baym2016spatiotemporal,
  title={Spatiotemporal microbial evolution on antibiotic landscapes},
  author={Baym, Michael and Lieberman, Tami D and Kelsic, Eric D and Chait, Remy and Gross, Rotem and Yelin, Idan and Kishony, Roy},
  journal={Science},
  volume={353},
  number={6304},
  pages={1147--1151},
  year={2016},
  publisher={American Association for the Advancement of Science}
}

@article{lambert2014bacteria,
  title={Bacteria and game theory: the rise and fall of cooperation in spatially heterogeneous environments},
  author={Lambert, Guillaume and Vyawahare, Saurabh and Austin, Robert H},
  journal={Interface focus},
  volume={4},
  number={4},
  pages={20140029},
  year={2014},
  publisher={The Royal Society}
}

@book{durrett2008probability,
  title={Probability models for DNA sequence evolution},
  author={Durrett, Richard},
  year={2008},
  publisher={Springer Science \& Business Media}
}

@article{yagoobi2021,
  author  = {Yagoobi, S. and Traulsen, A.},
  title   = {Fixation probabilities in network structured meta-populations},
  journal = {Scientific Reports},
  volume  = {11},
  number  = {1},
  pages   = {17979},
  year    = {2021}
}

@article{yagoobi2023,
  author  = {Yagoobi, S. and Sharma, N. and Traulsen, A.},
  title   = {Categorizing update mechanisms for graph-structured metapopulations},
  journal = {Journal of the Royal Society Interface},
  volume  = {20},
  number  = {200},
  pages   = {20220769},
  year    = {2023}
}

@article{durrett1994,
  author  = {Durrett, R. and Levin, S.},
  title   = {The importance of being discrete (and spatial)},
  journal = {Theoretical Population Biology},
  volume  = {46},
  number  = {3},
  pages   = {363--394},
  year    = {1994}
}

@article{durrett2014,
  author  = {Durrett, R.},
  title   = {Spatial evolutionary games with small selection coefficients},
  journal = {Electronic Journal of Probability},
  volume  = {19},
  number  = {121},
  pages   = {1--64},
  year    = {2014}
}

@article{coxdurrett2016,
  author  = {Cox, J. T. and Durrett, R.},
  title   = {Evolutionary games on the torus with weak selection},
  journal = {Stochastic Processes and their Applications},
  volume  = {126},
  number  = {8},
  pages   = {2388--2409},
  year    = {2016}
}

@article{cheng2026,
  author  = {Cheng, H. and Wang, H. and Meng, X.},
  title   = {Spatio-temporal evolution of cooperation: multistability, pattern formation, and chaos in resource-driven eco-evolutionary games},
  journal = {Journal of Mathematical Biology},
  volume  = {92},
  number  = {1},
  pages   = {12},
  year    = {2026}
}

@article{codling2008,
  author  = {Codling, E. A. and Plank, M. J. and Benhamou, S.},
  title   = {Random walk models in biology},
  journal = {Journal of the Royal Society Interface},
  volume  = {5},
  number  = {25},
  pages   = {813--834},
  year    = {2008}
}

@article{plank2025,
  author  = {Plank, M. J. and Simpson, M. J. and Baker, R. E.},
  title   = {Random walk models in the life sciences: including births, deaths and local interactions},
  journal = {Journal of the Royal Society Interface},
  volume  = {22},
  number  = {222},
  pages   = {20240422},
  year    = {2025}
}

@book{keener2021,
  author    = {Keener, J. P.},
  title     = {Biology in Time and Space: A Partial Differential Equation Modeling Approach},
  volume    = {50},
  publisher = {American Mathematical Society},
  year      = {2021}
}

@book{ewens2004mathematical,
  author = {Ewens, W.J.},
  publisher = {Springer},
  title = {{Mathematical population genetics}},
  year = 2004
}

@article{fisher1937wave,
  title={The wave of advance of advantageous genes},
  author={Fisher, Ronald Aylmer},
  journal={Annals of eugenics},
  volume={7},
  number={4},
  pages={355--369},
  year={1937},
  publisher={Wiley Online Library}
}

@article{kolmogorov1937etude,
  title={{\'E}tude de l’{\'e}quation de la diffusion avec croissance de la quantit{\'e} de mati{\`e}re et son application {\`a} un probl{\`e}me biologigue},
  author={Kolmogorov, Andrei},
  journal={Moscow Univ. Bull. Ser. Internat. Sect. A},
  volume={1},
  pages={1},
  year={1937}
}

@book{murray2002mathematical,
  title={Mathematical biology: I. An introduction},
  author={Murray, James D},
  volume={17},
  year={2002},
  publisher={Springer Science \& Business Media}
}

@article{van2003front,
  title={Front propagation into unstable states},
  author={Van Saarloos, Wim},
  journal={Physics reports},
  volume={386},
  number={2-6},
  pages={29--222},
  year={2003},
  publisher={Elsevier}
}

@article{hallatschek2007genetic,
  title={Genetic drift at expanding frontiers promotes gene segregation},
  author={Hallatschek, Oskar and Hersen, Pascal and Ramanathan, Sharad and Nelson, David R},
  journal={Proceedings of the National Academy of Sciences},
  volume={104},
  number={50},
  pages={19926--19930},
  year={2007},
  publisher={National Academy of Sciences}
}

@article{korolev2010genetic,
  title={Genetic demixing and evolution in linear stepping stone models},
  author={Korolev, Kirill S and Avlund, Mikkel and Hallatschek, Oskar and Nelson, David R},
  journal={Reviews of modern physics},
  volume={82},
  number={2},
  pages={1691--1718},
  year={2010},
  publisher={APS}
}

@article{houchmandzadeh2017fisher,
  title={Fisher waves: An individual-based stochastic model},
  author={Houchmandzadeh, Bahram and Vallade, Marcel},
  journal={Physical Review E},
  volume={96},
  number={1},
  pages={012414},
  year={2017},
  publisher={APS}
}

@article{kimura1962probability,
  title={On the probability of fixation of mutant genes in a population},
  author={Kimura, Motoo},
  journal={Genetics},
  volume={47},
  number={6},
  pages={713},
  year={1962}
}

@article{waclaw2015spatial,
  title={A spatial model predicts that dispersal and cell turnover limit intratumour heterogeneity},
  author={Waclaw, Bartlomiej and Bozic, Ivana and Pittman, Meredith E and Hruban, Ralph H and Vogelstein, Bert and Nowak, Martin A},
  journal={Nature},
  volume={525},
  number={7568},
  pages={261--264},
  year={2015},
  publisher={Nature Publishing Group UK London}
}

@article{noble2022spatial,
  title={Spatial structure governs the mode of tumour evolution},
  author={Noble, Robert and Burri, Dominik and Le Sueur, C{\'e}cile and Lemant, Jeanne and Viossat, Yannick and Kather, Jakob Nikolas and Beerenwinkel, Niko},
  journal={Nature ecology \& evolution},
  volume={6},
  number={2},
  pages={207--217},
  year={2022},
  publisher={Nature Publishing Group UK London}
}

@book{nowak2000virus,
  title={Virus dynamics: mathematical principles of immunology and virology: mathematical principles of immunology and virology},
  author={Nowak, Martin and May, Robert M},
  year={2000},
  publisher={Oxford University Press, UK}
}

@book{jackson1998classical,
  title     = {Classical Electrodynamics},
  author    = {Jackson, John David},
  year      = {1998},
  edition   = {3rd},
  publisher = {Wiley},
  address   = {New York}
}

@article{swanson2003virtual,
  title={Virtual and real brain tumors: using mathematical modeling to quantify glioma growth and invasion},
  author={Swanson, Kristin R and Bridge, Carly and Murray, JD and Alvord Jr, Ellsworth C},
  journal={Journal of the neurological sciences},
  volume={216},
  number={1},
  pages={1--10},
  year={2003},
  publisher={Elsevier}
}

@article{ferreira2002reaction,
  title={Reaction-diffusion model for the growth of avascular tumor},
  author={Ferreira Jr, SC and Martins, Marcelo Lobato and Vilela, MJ},
  journal={Physical Review E},
  volume={65},
  number={2},
  pages={021907},
  year={2002},
  publisher={APS}
}

@article{murray2003spatial,
  title={Spatial models and biomedical applications},
  author={Murray, James Dickson},
  journal={Mathematical Biology},
  year={2003},
  publisher={Springer}
}

@article{datta2013range,
  title={Range expansion promotes cooperation in an experimental microbial metapopulation},
  author={Datta, Manoshi Sen and Korolev, Kirill S and Cvijovic, Ivana and Dudley, Carmel and Gore, Jeff},
  journal={Proceedings of the National Academy of Sciences},
  volume={110},
  number={18},
  pages={7354--7359},
  year={2013},
  publisher={National Academy of Sciences}
}

@article{gandhi2016range,
  title={Range expansions transition from pulled to pushed waves as growth becomes more cooperative in an experimental microbial population},
  author={Gandhi, Saurabh R and Yurtsev, Eugene Anatoly and Korolev, Kirill S and Gore, Jeff},
  journal={Proceedings of the National Academy of Sciences},
  volume={113},
  number={25},
  pages={6922--6927},
  year={2016},
  publisher={National Academy of Sciences}
}

@article{maslovskaya5388371allen,
  title={The Allen-Cahn-Based Approach to Cross-Scale Modeling Bacterial Growth Controlled by Quorum Sensing},
  author={Maslovskaya, Anna and Shuai, Yixuan and Kuttler, Christina},
  journal={Available at SSRN 5388371}
}

@inproceedings{levin2012mathematical,
  title={Mathematical and computational challenges in the study of complex adaptive microbial systems},
  author={Levin, Simon A and Bonachela, JA and Nadell, CD},
  booktitle={The Social Biology of Microbial Communities: Workshop Summary},
  year={2012},
  organization={National Academies Press}
}

@book{gardiner2004handbook,
  title={Handbook of stochastic methods},
  author={Gardiner, Crispin W and others},
  volume={3},
  year={2004},
  publisher={springer Berlin}
}

@article{Greulich2012,
  title = {Mutational Pathway Determines Whether Drug Gradients Accelerate Evolution of Drug-Resistant Cells},
  author = {Greulich, Philip and Waclaw, Bartlomiej and Allen, Rosalind J.},
  journal = {Physical Review Letters},
  volume = {109},
  number = {8},
  pages = {088101},
  year = {2012},
  doi = {10.1103/PhysRevLett.109.088101}
}

@article{Hermsen2012,
  title = {On the Rapidity of Antibiotic Resistance Evolution Facilitated by a Concentration Gradient},
  author = {Hermsen, Rutger and Deris, John B. and Hwa, Terence},
  journal = {Proceedings of the National Academy of Sciences},
  volume = {109},
  number = {27},
  pages = {10775--10780},
  year = {2012},
  doi = {10.1073/pnas.1117716109}
}

@article{Greulich2015,
  title = {Growth-Dependent Bacterial Susceptibility to Ribosome-Targeting Antibiotics},
  author = {Greulich, Philip and Scott, Matthew and Evans, Martin R. and Allen, Rosalind J.},
  journal = {Molecular Systems Biology},
  volume = {11},
  number = {3},
  pages = {796},
  year = {2015},
  doi = {10.15252/msb.20145949}
}

@article{Meredith2015,
  title = {Bacterial Temporal Dynamics Enable Optimal Design of Antibiotic Treatment},
  author = {Meredith, Hannah R. and Lopatkin, Allison J. and Anderson, Deverick J. and You, Lingchong},
  journal = {PLoS Computational Biology},
  volume = {11},
  number = {4},
  pages = {e1004201},
  year = {2015},
  doi = {10.1371/journal.pcbi.1004201}
}

@article{Coates2018,
  title = {Antibiotic-Induced Population Fluctuations and Stochastic Clearance of Bacteria},
  author = {Coates, Jessica and Park, Bo Ryoung and Le, Dai and Simsek, Emrah and Chaudhry, Waqas and Kim, Minsu},
  journal = {eLife},
  volume = {7},
  pages = {e32976},
  year = {2018},
  doi = {10.7554/eLife.32976}
}

@article{Baquero2021,
  title = {Proximate and Ultimate Causes of the Bactericidal Action of Antibiotics},
  author = {Baquero, Fernando and Levin, Bruce R.},
  journal = {Nature Reviews Microbiology},
  volume = {19},
  pages = {123--132},
  year = {2021},
  doi = {10.1038/s41579-020-00443-1}
}

@article{ElHachem2019,
  title = {Revisiting the Fisher--Kolmogorov--Petrovsky--Piskunov Equation to Interpret the Spreading--Extinction Dichotomy},
  author = {El-Hachem, M. and McCue, S. W. and Simpson, M. J.},
  journal = {Proceedings of the Royal Society A},
  volume = {475},
  number = {2229},
  pages = {20190378},
  year = {2019},
  doi = {10.1098/rspa.2019.0378}
}

@article{Gorgi2026GeometricOrdering,
  author = {Gorgi, Melika and Kasallis, S. J. and Trinh, C. and Ortiz de Ora, L. and Wiles, T. J. and Siryaporn, Albert},
  title = {Geometric ordering in bacterial communities},
  journal = {Proceedings of the National Academy of Sciences},
  year = {2026},
  volume = {123},
  number = {20},
  pages = {e2526643123},
  doi = {10.1073/pnas.2526643123}
}

@misc{MoranDynamicsGitHub,
  author = {Gorgi, Melika},
  title = {MoranDynamics},
  year = {2026},
  publisher = {GitHub},
  journal = {GitHub repository},
  howpublished = {\url{https://github.com/melikagorgi/MoranDynamics}}
}

\end{document}